\documentclass[twocolumn]{nature-Fig}
\usepackage{color}
\usepackage{soul}

\pdfoutput=1

\usepackage{graphicx}
\usepackage{amsmath, amssymb, amsfonts,bm}
\usepackage{latexsym}
\usepackage{bbm}
\usepackage[protrusion=true, expansion=true]{microtype}

\title{Coherence and multimode correlations from vacuum fluctuations in a microwave superconducting cavity}

\author{Pasi L\"ahteenm\"aki$^{1}$, Gheorghe Sorin Paraoanu$^1$, Juha Hassel$^{2}$, \& Pertti ~J. Hakonen$^1\footnote{Correspondence should be addressed to pertti.hakonen@aalto.fi}$}

\begin{document}

\maketitle
\begin{affiliations}
\item Low Temperature Laboratory, Department of Applied Physics, Aalto University School of Science, P.O. Box 15100, FI-00076 AALTO, Finland
\item VTT Technical Research Centre of Finland, FI-02044, Espoo, Finland
\end{affiliations}

\date{\today}

\begin{abstract}

The existence of vacuum fluctuations is one of the most important predictions of modern quantum field theory. In the vacuum state, fluctuations occurring at different frequencies are uncorrelated. However, if a parameter in the Lagrangian of the field is modulated by an external pump, vacuum fluctuations stimulate spontaneous downconversion processes, creating squeezing between modes symmetric with respect to half of the frequency of the pump.

Here we show that by double parametric pumping of a superconducting microwave
cavity, it is possible to generate another fundamental type of correlation, namely coherence between photons in separate frequency modes. The coherence correlations are tunable by the phases of the pumps and are established by a quantum fluctuation that stimulates the simultaneous creation of two photon pairs. Our analysis indicates that the origin of this vacuum-induced coherence is the absence of “which-way” information in the frequency space.

\end{abstract}

\clearpage

\section{INTRODUCTION}

A direct consequence of the Heisenberg uncertainty principle is that quantum fields,
even in the vacuum state, are teeming with fluctuations. Modern quantum field theory predicts that these fluctuations are not only a useful mathematical representation, but they can produce observable effects, encompassing vastly different physical scales -  from atoms to black holes.
Among these are the Purcell effect\cite{purcelle}, the Lamb shift of the atomic states\cite{lamb}, the Schwinger effect\cite{schwinger}, the Hawking radiation\cite{hawking}, and the Casimir effect\cite{casimir}.

Recently, in parallel to the interest in fundamental physics\cite{ussC}, a novel approach has emerged - the idea of engineering the quantum vacuum to create novel devices and protocols for quantum technologies\cite{carloss}. Parametrically modulated superconducting circuits have attracted significant interest\cite{ussA,ussB,nori2011}, motivated also by the demand for parametric amplifiers where the added noise is pushed to the quantum limit\cite{dpa2009,clerk2010,champara,yales,eichlers}. In these systems, it has been demonstrated that vacuum fluctuations present at the input port trigger the creation of real microwave photons\cite{yurke1988,castellanos2008,mallet2011,eichler2011,wilson2011,pasi2013,menzel2012}. Some of these experiments provide beautiful analogies with the motion of a mirror in free space\cite{wilson2011,pasi2013,ussD,revmodph}, also referred to as dynamical Casimir effect. For the development of quantum computing with continuous variables (CV) in microwave circuits\cite{chalmers-theory}, these results demonstrate that a key operation is now available experimentally: two-mode squeezing\cite{yuen1976,caves1985} of the modes $a$ and $b$, leading to a nonzero correlation $\langle ab\rangle \neq 0$.
However,
there exists another fundamental type of correlation between microwave fields, yielding $\langle a^{\dag} b\rangle \neq 0$, which is instrumental in CV quantum computing.
This is a measure of coherence of the fields in the two modes, and it appears ubiquitously in the modeling of beam-splitter and interference phenomena. From the perspective of quantum vacuum engineering, achieving this correlation is problematic: vacuum fluctuations at different times and frequencies are completely uncorrelated. This is such a precise feature, that vacuum fluctuations can even serve as
perfect random number generators\cite{qrng}.
Even when the fluctuations trigger the downconversion of a higher energy photon into two photons in the modes $a$ and $b$, as in the case of a single-pump dynamical Casimir effect, the correlation $\langle a^{\dag} b\rangle$ remains zero.

Here we demonstrate that two-mode coherence correlations can be obtained from the dynamical Casimir effect by employing another fundamental quantum-mechanical principle, namely the absence of which-way information\cite{mandel1991,scully2000}, which is applied here in the frequency space. Specifically, we study the parametric modulation of the quantum vacuum under the action of two pumps. We demonstrate that a non-zero coherent correlation $\langle \tilde{a}_1^\dagger \tilde{a}_2 \rangle$ is established between two frequency modes $\tilde{a}_1$ and $\tilde{a}_2$, through downconversion processes in the two pumps triggered by the same vacuum fluctuation in a third frequency mode $\tilde{a}_0$. This correlation is non-zero provided that it is not possible to specify from which pump the down-converted photons at mode $\tilde{a}_0$ originate from. Additionally, due to this coherence effect, a highly-populated “bright” mode and, orthogonal to it, a zero-population “dark” mode can be defined in the subspace of the two frequency modes. Dark states appear as a result of destructive quantum interference.
Typically, in atomic physics, optics, and more recently in superconducting microwave systems, dark states appear in Rabi-driven multilevel quantum systems\cite{pappas2010,arimondo1996,vuleti2011,agarwal2013,semba2014,kumar2015}, or, in the case of resonators, through sideband driving\cite{marquardt2007,zakka2011}, both of which require Hamiltonians of the type $a^\dagger b + h.c$. to simply transfer population from one level to another. In contrast, here our Hamiltonian is of parametric type $a^\dagger a^\dagger + h.c.$, which creates populations from the vacuum instead of transferring quanta between modes. We also show that this parametric coherence effect is a phase-sensitive phenomenon and, therefore, it can be controlled by the relative phase of the applied pumps. Our result opens a way for realizing highly complex multi-mode entangled states displaying both coherence and squeezing correlations with CV. In terms of the modes $\tilde{a}_0$, $\tilde{a}_1$, and $\tilde{a}_2$, the state we created is a CV tripartite state\cite{giedke, adesso} of bisymmetric type\cite{serafini, gadesso}, which at low powers, in the subspace with at most two photons, becomes a W-type state\cite{dur}. Tripartite bisymmetric states are a resource for protocols such as one-to-two telecloning\cite{loock2001,adesso2007}, while W states can be used for fundamental tests of quantum mechanics\cite{realism,thispaper}. Our approach can be generalized in a straightforward way to multiple modes and pumps, opening the way to the implementation of algorithms such as bosonic sampling\cite{aaronson} and the realization of CV cluster states for one-way quantum computing\cite{pfister2014,hans} in the microwave regime.

\begin{figure}
\vspace{0.3cm}
\centering
\includegraphics[width=0.4\linewidth]{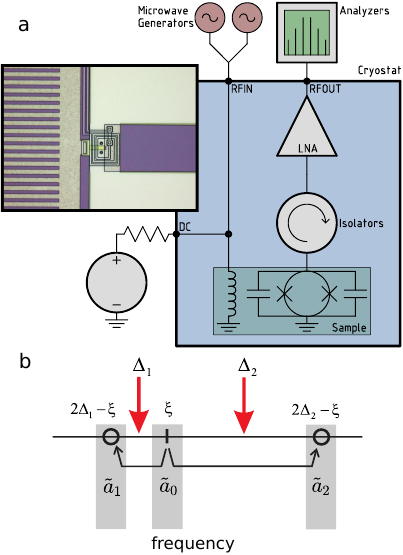}
\caption{{\bf Overview of the measurement setup and pumping.} (a) Simplified schematic of the measurement setup (see Supplementary Figs. 3 and 4 from Supplementary Note 7 for more detailed schematics). The investigated resonator was made using a capacitively shunted superconducting quantum interference device (SQUID), fabricated using a NbAlO$_x$Nb trilayer process at VTT\cite{seppaVTT}. The device forms a flux tunable resonator which acts as a tunable boundary condition for the reflected waves in the attached 50 $\Omega$ transmission line\cite{pasi2013}. The resonance is tuned down from the maximum value (nominally 10 GHz) to 5 GHz using dc-flux biasing. The designed critical current for the junctions of the SQUID was $I_C = 33 {\rm \ \mu A}$, and the total parallel capacitance was $C = 40{\ \rm pF}$. We estimate the effective temperature to remain below 35 mK at the highest pumping powers used in these experiments while the cryostat base temperature is held at 15 mK. This corresponds to thermal occupation number $\overline{n}=10^{-3}$. Modulation of the flux through the SQUID is realized through a lithographically fabricated rectangular spiral coil underneath the junction layer. (b) Illustration of the first reflections of a frequency $\xi=\omega -\omega_{\rm{res}}$ with respect to half the pump frequencies. The short-hand notation of modes $\tilde{a}_0$, $\tilde{a}_1$, and $\tilde{a}_2$ signifies the cavity output modes  $\tilde{a}_{\rm out}[\xi ] $, $\tilde{a}_{\rm out}[2\Delta_{1}- \xi] $, and $\tilde{a}_{\rm out}[2\Delta_{2}- \xi ] $, respectively.}

\label{fig_schem}
\end{figure}

\section{RESULTS}
{\bf Theoretical predictions}\ \\
Our system consists of a superconducting resonator (see Fig. \ref{fig_schem}) which can be described by the effective Hamiltonian\cite{pasi2013}
\small
\begin{equation}
H = \hbar\omega_{\rm res} a^{\dag}a + \frac{\hbar}{2i}\sum_{p=1,2}\left[\alpha_{p}^{*}e^{i\omega_{p}t+\varphi_p} -\alpha_{p}e^{-i\omega_{p}t-\varphi_p}\right](a + a^{\dag})^{2},
\end{equation}
\normalsize
where $\alpha_{p}$ is the strength of pump $p$ coupled into the resonator, in units of frequency, and $\varphi_p$ specifies the pump phase. Denoting the decay rate of the resonator by $\kappa$ and setting $\xi=\omega -\omega_{\rm{res}}$, the output field $\tilde{a}_{\rm out}[\xi ] = \tilde{a}_{\rm in}[\xi ] + \sqrt{\kappa}\tilde{a}[\xi ]$ is obtained within the input-output formalism by solving iteratively the Heisenberg-Langevin equations
in a rotating frame at the resonator frequency defined by $\tilde{a} (t) = e^{i \omega_{\rm res} t} a(t)$.
We obtain a nested structure

\small
\begin{multline}\label{finalsolution}
\tilde{a}_{\rm out} [\xi ] = \left[1-\frac{\kappa\chi (\xi )}{1 -
\sum_{p=1}^{2}{\cal N}^{[p]}_{1} (\xi )}\right]\tilde{a}_{\rm in}[\xi ] -\\
\frac{\kappa\chi (\xi )}{1 -
\sum_{p=1}^{2}{\cal N}^{[p]}_{1} (\xi )}\sum_{p=1}^{2} \frac{\alpha_{p}\chi^{*}(2\Delta_{p} - \xi)}{{\cal N}^{[p]}_{2} (\xi )}
\left[(\tilde{a}_{\rm in}[2\Delta_{p} - \xi])^{\dag}+
\frac{\alpha_{\bar{p}}^{*}\chi(2\Delta_{\bar{p}} - 2\Delta_{p}+ \xi)}{{\cal N}^{[p]}_{3} (\xi )}
\tilde{a}_{\rm in}[2\Delta_{\bar{p}} - 2\Delta_{p}+ \xi]
+ \dots \right].
\end{multline}
\normalsize
Here $\chi(\xi) = (\kappa/2-i\xi)^{-1}$ is the electrical susceptibility of the resonator,  $\bar{p} \in \{1,2\}\setminus \{p\}$, ${\cal N}^{[p]}_{j}$ denote normalization factors (see the Supplementary Notes 1-3), and $\Delta_p=\omega_p/2-\omega_{\rm res}$. %

For simplification, we truncate Eq. (\ref{finalsolution}) to the first-order reflections of $\xi$ with respect to the pumps, and assume $|\Delta_{p} |>> |\xi |$.
Higher-order reflection (Supplementary Fig. 1) amplitudes are suppressed by a factor proportional to the cavity susceptibility, however they can be observed experimentally at high enough pumping power (Supplementary Fig. 6). In this first-order approximation we have ${\cal N}^{[p]}_{2} (\xi ) \approx 1$, and we use a pump power parametrization by a squeezing parameter $\lambda$ and power asymmetry angle $\theta$, $\alpha_1 \chi^{*}(2\Delta_1)= \cos\theta \tanh (\lambda /2 )  e^{i\varphi_1}$, and $\alpha_2 \chi^{*}(2\Delta_2)=  \sin\theta \tanh (\lambda /2)  e^{i\varphi_2}$. This allows us to calculate the correlations in the output field of the cavity $\tilde{a}_{out}$, with the vacuum as the input state. We find that each pump produces squeezing correlations $\left< \tilde{a}_{\rm out}[\xi ]  \tilde{a}_{\rm out}[2\Delta_{1}-\xi '] \right> =  \frac{1}{2}\cos\theta \exp(i\varphi_1 )\sinh 2\lambda \times \delta (\xi -\xi ')$, $\left< \tilde{a}_{\rm out}[\xi ]  \tilde{a}_{\rm out}[2\Delta_{2}-\xi '] \right> =  \frac{1}{2}\exp(i\varphi_2 )\sin\theta \sinh 2\lambda \times \delta (\xi -\xi ')$ as well as noise at $2\Delta_1 -\xi$ and $2\Delta_2 -\xi$, according to the dynamical Casimir effect power noise formula\cite{pasi2013} for each pump separately,
$\left< (\tilde{a}_{\rm out}[2\Delta_{1}-\xi ])^{\dag}  \tilde{a}_{\rm out}[2\Delta_{1}-\xi '] \right> =  \cos^{2}\theta \sinh^{2}\lambda \times \delta (\xi -\xi ')$,
$\left< (\tilde{a}_{\rm out}[2\Delta_{2}-\xi ])^{\dag}  \tilde{a}_{\rm out}[2\Delta_{2}-\xi '] \right> =  \sin^{2}\theta \sinh^{2}\lambda \times \delta (\xi -\xi ')$, while at
$\xi$ the noise is produced by the action of both pumps,
$\left< (\tilde{a}_{\rm out}[\xi ])^{\dag}  \tilde{a}_{\rm out}[\xi '] \right> =  \sinh^{2}\lambda \times \delta (\xi -\xi ')$.
Surprisingly, the simultaneous action of the two pumps produces the correlation $\left< (\tilde{a}_{\rm out}[2\Delta_{1}-\xi ])^{\dag}  \tilde{a}_{\rm out}[2\Delta_{2}-\xi '] \right> =  \left(\left< (\tilde{a}_{\rm out}[2\Delta_{2}-\xi ])^{\dag}  \tilde{a}_{\rm out}[2\Delta_{1}-\xi '] \right> \right)^{*}$ between the frequencies $2\Delta_1 -\xi$ and $2\Delta_2 -\xi$,
\small
\begin{equation}
\left< (\tilde{a}_{\rm out}[2\Delta_{1}-\xi ])^{\dag}  \tilde{a}_{\rm out}[2\Delta_{2}-\xi '] \right> =
\frac{\sin 2\theta }{2} e^{i(\varphi_{2}-\varphi_{1})}\sinh^2\lambda   \times \delta (\xi -\xi '),
\end{equation}
\normalsize
which can be regarded as a coherence between the extremal modes $2\Delta_1 -\xi$ and $2\Delta_2 -\xi$, mediated by the middle vacuum fluctuation. {From the expression of the correlations above we can construct the covariance matrix, which has a bisymmetric structure\cite{serafini, gadesso}.} {Note that, since the Hamiltonian is quadratic in the field operators, the state is Gaussian. Therefore, the first-order correlations presented here provide a complete characterization of the output field.}

Based on the structure of Eq (2) we can introduce `bright' and `dark' modes,
\small
\begin{eqnarray}
\tilde{b}[\xi ] &=& \cos\theta e^{-i\varphi_1} \tilde{a}[2\Delta_{1} - \xi] + \sin\theta  e^{-i\varphi_2}\tilde{a}[2\Delta_{2} - \xi] \label{bfield},\\
\tilde{d}[\xi ] &=& \sin\theta e^{-i\varphi_1} \tilde{a}[2\Delta_{1} - \xi] - \cos\theta  e^{-i\varphi_2}\tilde{a}[2\Delta_{2} - \xi]\label{afield},
\end{eqnarray}
\normalsize
and similar relations for the output and input modes. These two modes are orthogonal to each other, spanning the Hilbert space of the extremal modes $2\Delta_{1} - \xi$ and $2\Delta_{2} - \xi$.
{Experimentally, the asymmetry angle $\theta$ in the bright and dark modes is
determined from measurements of the noise power generated by each pump through the dynamical Casimir effect.}
The definition of the modes $\tilde{b}$ and $\tilde{d}$ resembles a beam-splitter/merging operation in frequency (rather than in space, as in usual interferometers), where a mode is separated into two branches with a distance $2 (\Delta_{1} -\Delta_{2})$ between them. {Indeed, when the modes $\tilde{a}[2\Delta_1-\xi]$ and $\tilde{a}[2\Delta_2-\xi]$ are rotated by some angles $\varphi_1$ and $\varphi_2$ and combined as in Eqs. (4,5), this vacuum-induced coherence manifests as an interference effect, producing the extinction of power in the dark mode and the maximization of power in the bright mode. Specifically, for the correlations involving the dark and bright modes we obtain }
\small
\begin{eqnarray}
\left< (\tilde{b}_{\rm out}[\xi ])^{\dag}  \tilde{b}_{\rm out}[\xi '] \right> &=&  \sinh^2\lambda \times \delta (\xi -\xi '),\\
\left< \tilde{a}_{\rm out}[\xi ]  \tilde{b}_{\rm out}[\xi '] \right> &=&  \frac{1}{2}\sinh 2 \lambda \times \delta (\xi -\xi '), \\
\left< \tilde{a}_{\rm out}[\xi ]  \tilde{d}_{\rm out}[\xi '] \right> &=&\left<\tilde{b}_{\rm out}[\xi ]  \tilde{d}_{\rm out}[\xi '] \right> = 0 ,\\
\left< (\tilde{d}_{\rm out}[\xi ])^{\dag}  \tilde{d}_{\rm out}[\xi '] \right> &=&  0.
\end{eqnarray}
\normalsize

These correlations reflect the structure of the output state of the resonator  $|{\rm vac}\rangle_{\rm out}$  which fulfills $\tilde{a}_{\rm out} |{\rm vac}\rangle_{\rm out}=0$. For the truncated form of Eq. (\ref{finalsolution}), the output state is a two-mode squeezed state in terms of the operators $\tilde{a}[\xi ]$ and $\tilde{b}[\xi ]$, $|{\rm vac}\rangle_{\rm out} = S^\dagger|{\rm vac}\rangle_{\rm in}$, where $S =  \exp\left[ \lambda \tilde{a}_{\rm in} \tilde{b}_{\rm in} - h.c. \right]$. {In the subspace containing at most two excitations, the tripartite state is in the W class, see Supplementary Note 5.}

The disappearance of power in the dark mode is a particular manifestation of coherence which has been employed for many applications, such as quantum memory\cite{semba2014} and STIRAP in superconducting qubit systems\cite{kumar2015}. Note that in the usual type of dark state, the destructive interference is produced at the same frequency, for example, on a quantum state of a qubit or of a mode of a field. Here, the two modes at $2\Delta_1 - \xi$ and $2\Delta_2 - \xi$ are separated in frequency and do not overlap\cite{zeilinger2009}. The destructive interference effect is created by the vacuum fluctuations at $\xi$ which trigger two correlated two-photon parametric downconversion processes in the pumps.

As for standard quantum interference, the lack of path information\cite{zeilinger2014,review-delayed} is critical for the interference between extremal frequencies. Here, instead of which-path information in space, we deal with absence of which-color information\cite{mandel1991,scully2000} : for a real photon at frequency $\xi$, there is no way of knowing from which of the two spontaneous parametric downconversion processes it came from {(see Supplementary Note 4).}

\begin{figure}
\centering
\includegraphics[width=0.99\linewidth]{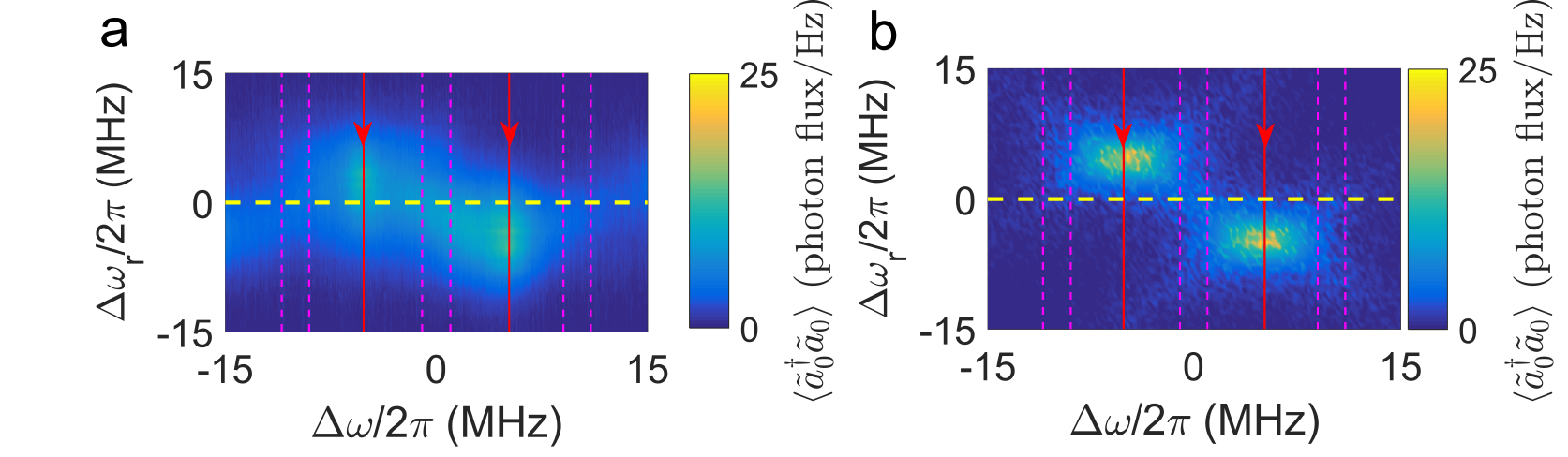}
{
\caption{{\bf Noise power spectra.} (a) Measured noise power spectra corresponding to the correlator $\left< (\tilde{a}_{\rm out}[\xi ])^{\dag}  \tilde{a}_{\rm out}[\xi ] \right>$ for fixed half pump frequencies $\omega_1/2 = (2\pi)\times4.995$ GHz, $\omega_2/2 = (2\pi)\times5.005$ GHz at different values of the resonance frequency $\Delta\omega_{\rm r}/2\pi$ and measurement frequency $\Delta\omega/2\pi$ relative to $(\omega_1/2 +\omega_2/2)/2 = (2\pi)\times5$ GHz for a pump power corresponding to 20 photon flux/Hz emerging from the cavity at the center frequency of 5 GHz.
The colorbar is scaled in dB relative to vacuum. The dashed magenta lines indicate the measurement bands within which the correlation data were collected. The red lines indicate the locations of the half of the pump frequencies $\omega_1$ and $\omega_2$.
The dashed yellow line indicates the symmetry point which was used for measuring the data depicted in Fig. \ref{fig_mim}. The external and internal cavity decay rates for this sample were $\kappa_E = 2\kappa_I \approx {\rm 36\ MHz} $.
(b) Noise power as predicted by Eq. (\ref{finalsolution}) including only first-order reflections with $\kappa=\kappa_E+\kappa_I$ taken from the measurement (Supplementary Fig. 2).}
}
\label{fig_spow}
\end{figure}
\ \\
{\bf Experimental results for correlations}\ \\
To test the above predictions, we pumped our sample using two phase-locked microwave generators at frequencies 9.99 GHz and 10.01 GHz. The output field $\tilde{a}_{\rm out}$ from the resonator is amplified, yielding a signal that is further downconverted and digitized. The Fourier-transformed amplified field is measured in a bandwidth around the three frequencies of interest. The resulting fields are denoted by $\tilde{a}_{0}$, $\tilde{a}_{1}$, $\tilde{a}_{2}$ (see Methods), corresponding to the output fields $\tilde{a}_{\rm out}[\xi ]$, $\tilde{a}_{\rm out}[2\Delta_{1} -\xi]$, and $\tilde{a}_{\rm out}[2\Delta_{2} -\xi]$ respectively. Upon substraction of the added noise of the amplifiers, the correlations between the amplified fields provide a direct measurement of the corresponding correlations of the resonator output field\cite{eichler2011,DaSilva2010}. {For additional information see Supplementary Notes 7-8.}
The measured noise power data can be reproduced by simulations using the theoretical results presented earlier. The data are presented in Fig. \ref{fig_spow} a), while the corresponding numerical result from Eq. (\ref{finalsolution}) is presented  in Fig. \ref{fig_spow} b). {The measured correlations, as well as the analytical and the simulation results are normalized with respect to the vacuum state values, corresponding to $\lambda = 0$. Thus, the correlations at finite pumping $\lambda \neq 0$ are expressed in dB (with vacuum as reference). The noise power levels can also be expressed in absolute units (photon flux/Hz).

} For symmetry reasons, we chose to analyze correlations when the microwave resonance lies at half of the average of the two pump frequencies, that is $\Delta_{1}\approx -\Delta_{2}$, in which case the tripartite correlations are strongest.

\begin{figure}
\centering
\includegraphics[width=0.9\linewidth]{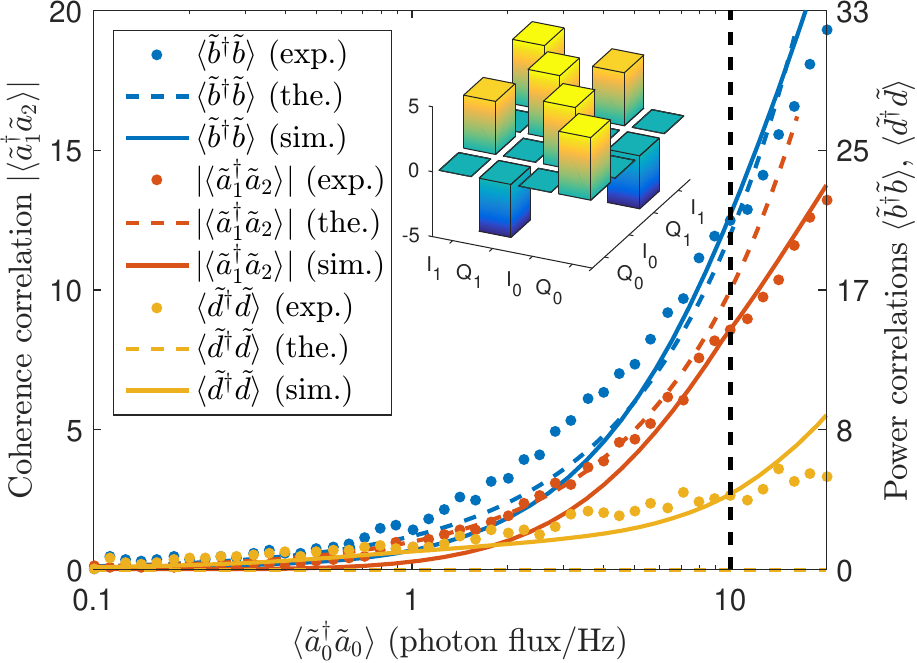}
{
\caption{{\bf Correlations: experimental data and theoretical predictions.} We present the measured correlators (specified in the inset label) vs. photon flux/Hz at the output of the cavity measured at the center frequency of 5 GHz; see text for the relation between the photon flux and the squeezing parameter. The symbols refer to measured data, while the curves are predictions from theory (dashed) and simulations (solid). Our simulation includes the higher order reflections and it reproduces also the residual population in the dark mode. The 3-D histogram represents the measured covariance matrix corresponding to the fields $\tilde{a}_0$ and $\tilde{a}_1$, where the mode quadratures defined as $I_0 = \frac{1}{2}(\tilde{a}_0+\tilde{a}_0^\dagger)$, $Q_0 = \frac{1}{2i}(\tilde{a}_0-\tilde{a}_0^\dagger)$, $I_1 = \frac{1}{2}(\tilde{a}_1+\tilde{a}_1^\dagger)$ and $Q_1 = \frac{1}{2i}(\tilde{a}_1-\tilde{a}_1^\dagger)$, yield two-mode squeezing correlations $|\langle \tilde{a}_0 \tilde{a}_1 \rangle|$ across the first pump. For additional correlations see Supplementary Fig. 5.}\label{fig_mim}
}
\end{figure}

Fig. \ref{fig_mim} displays the most relevant correlators determined as  a function of the number of photons in the resonator. All these correlators can be calculated from the measured field quadratures of the output field. {The simulation is based on the Langevin equation corresponding to the Hamiltonian Eq. (1) (see also Eq. (6) in Supplementary Note 1), with white noise as input, and the theoretical curves are obtained from Eqs. (3,6,9). We find that the behavior of these correlators agrees well with the theory - squeezing correlations exist between neighboring frequencies and vacuum-induced coherence correlations between the extremal ones. Indeed, the correlators grow nearly exponentially with the squeezing parameter, as predicted by theory. Furthermore, the ratio between various correlators is close to the expectations obtained from Eq. (\ref{finalsolution}), with only the first-order pump reflections included.}

From Fig. \ref{fig_mim} we note that the coherence is preserved also in the limit of small number of photons. This shows that our result is fundamentally different from tripartite slit-interferometer schemes\cite{haroche}, where the coherence between two modes is obtained only in the limit of a large number of photons in the pump mode.

\begin{figure}
\vspace{0.5cm}
\centering
\includegraphics[width=0.99\linewidth]{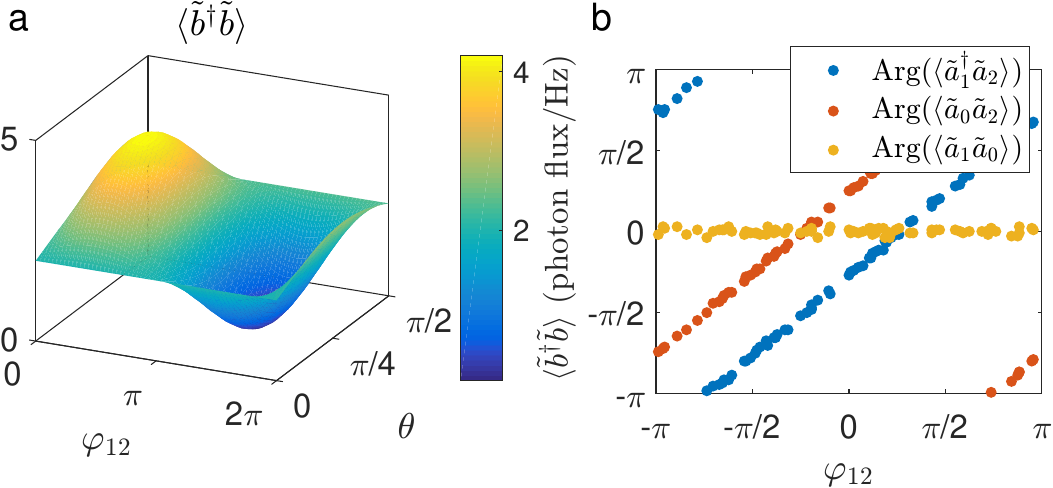}
\caption{{\bf Phase sensitivity of correlations.} (a) Bright/dark mode amplitude vs. phase difference between the pumps $\varphi_{12}= \varphi_{1}-\varphi_{2}$ and pump amplitude asymmetry parameter $\theta$ (symmetric pumps have $\theta = \pi/4$). The maximum value corresponds to the bright mode amplitude while the minimum refers to the dark mode. (b) Measured argument of the complex quadrature correlations vs. $\varphi_{12}$, where $\varphi_2$ is varied and $\theta \approx \pi/4$.}
\label{figbd}
\end{figure}

{The phase-dependent interplay between the bright and dark modes is presented in Fig. \ref{figbd}a). By adjusting the phase difference $\varphi_{12} = \varphi_1-\varphi_2$ of the pumps, the measured correlator power will shift from $\langle\tilde{b}^\dagger \tilde{b}\rangle$ into $\langle\tilde{d}^\dagger\tilde{d}\rangle$ as seen in Fig. \ref{figbd}a. The division of power between the dark and bright modes depends also on the asymmetry of the two pump amplitudes, \emph{ i.e.} the parameter $\theta$. Thus, we find $\theta$ by measuring the noise power generated by each pump separately.} Fig. \ref{figbd}b illustrates the change in the argument of the correlator when the phase of one pump is varied relative to the other one. {We note that the possibility of applying phase shifts on the modes by the rotation of the pump phases is a key operation in CV quantum computing\cite{loock2005}.}

\begin{figure}
\centering
\includegraphics[width=0.7\linewidth]{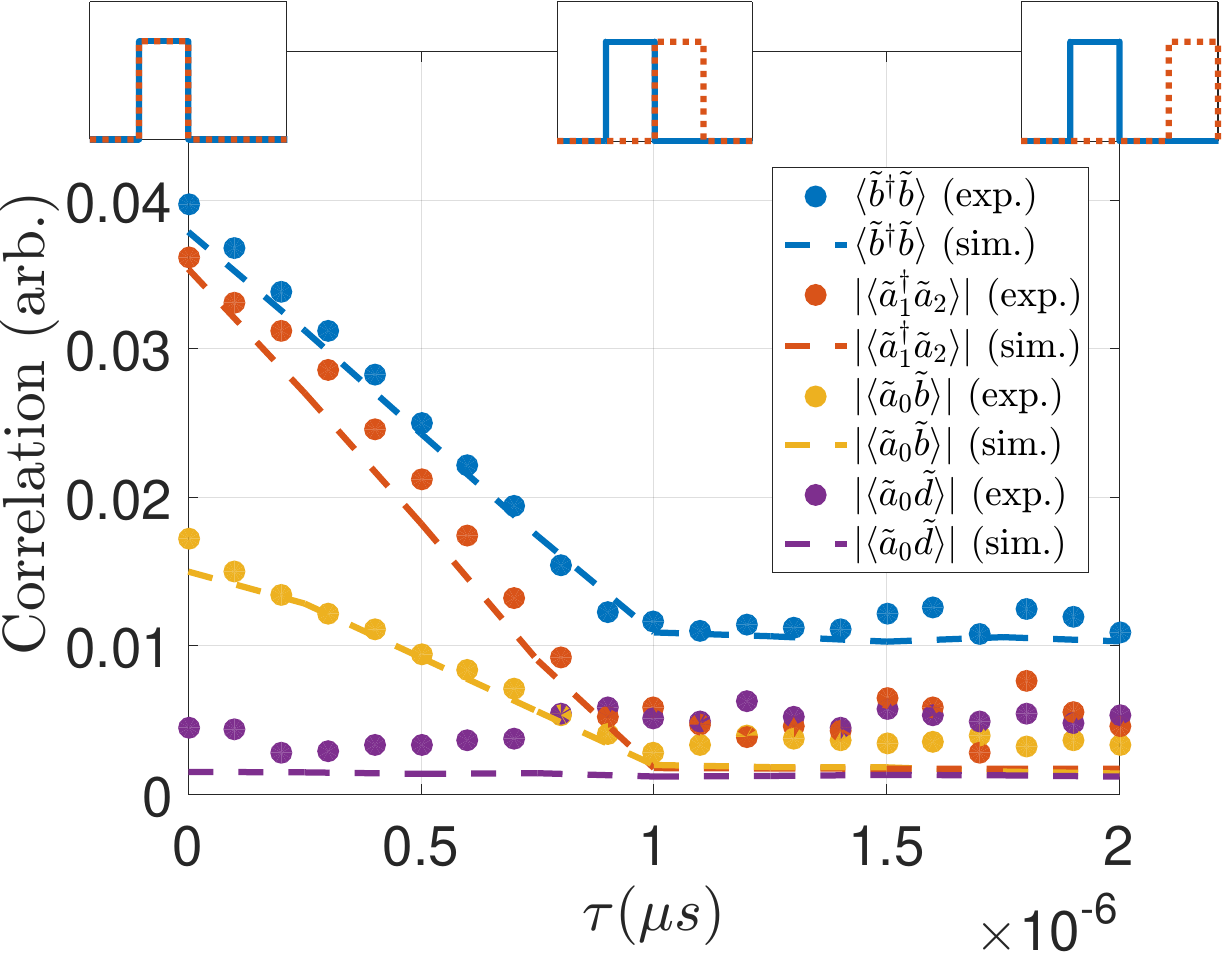}
{
\caption{{\bf Pulsed-pump measurements.} Decrease of vacuum-induced coherence $|\langle\tilde{a}_1^\dagger\tilde{a}_2\rangle|$ when the overlap time $\tau$ of the parametric pump pulses is varied. The coherence decreases linearly (see Supplementary Note 6) as the pumps become non-overlapping. Complete vanishing of the correlation between the extremal frequencies with zero overlap of the pulses implies that the photons are created simultaneously. For this experiment the sample was replaced with a new higher $Q$ chip with $\kappa_I \approx {\rm 1\ MHz}$, $\kappa_E \approx {\rm 2\ MHz}$. The pulse width for both pump tones was $1 {\rm\ \mu s}$. Three other correlators specified in the inset labeling are also depicted. %
Dashed lines are from time domain simulations based on Eq. (6) from the Supplementary Note 1 and the dots are experimental data. The difference in the values of $\langle \tilde{b}^\dagger \tilde{b}\rangle$ and $\langle \tilde{d}^\dagger \tilde{d}\rangle$ when the pulses do not overlap is due to the asymmetric power generation by the dynamical Casimir effect in each of the pumps separately.}\label{figpulse}
}
\end{figure}

\ \\
{\bf Correlations in time-domain}\ \\
{The creation of vacuum-induced coherence was investigated in pulsed pump tone experiments, in which the overlap of the pump signals was varied as illustrated in Fig. \ref{figpulse}. The results show that, in order to obtain a non-zero coherence correlation, the downconversion processes have to occur simultaneously.} We observe that the coherence $|\langle\tilde{a}_1^\dagger \tilde{a}_2\rangle|$ is reduced proportionally to the decrease of pulse overlap, while the squeezing correlations $|\langle\tilde{a}_1 \tilde{a}_0\rangle|$ and $|\langle\tilde{a}_0 \tilde{a}_2\rangle|$ remain. {The linear dependence of the coherence correlation on the overlap can be obtained through a fully analytical relation (see Supplementary Note 6). This demonstrates that the generation of the coherence requires overlapping pump signals and simultaneous creation of photons, during which the which-color information is not available. When the pump pulses are separated in time, the information about which pump has generated the photons becomes accessible in principle, and the coherence is suppressed.}

\section{DISCUSSION}

Our work on tripartite microwave correlations can be extended towards multipartite entangled states\cite{pfister2014}, which would yield a platform for universal quantum computation using CV, as recently proposed in Ref.\cite{chalmers-theory}. For example, the creation of cluster states in microwave cavities would be an important step towards realising a superconducting one-way computer. Such multipartite entangled states require pulsed microwave pumping for entangling different frequencies, a scheme that has been shown to be a fully functioning concept in our work.

\section{ACKNOWLEDGEMENTS}

We thank K. M{\o}lmer, Sir P. Knight, Y. Makhlin, C. Sab\'in, H. Briegel, W. D\"ur, D. Bruschi, G. Adesso, and A. Simonsen for discussions and M. Kiviranta and L. Gr\"onberg for contributions in fabrication. This work used the cryogenic facilities of the Low Temperature Laboratory infrastructure at OtaNano, Aalto University. We acknowledge financial support from FQXi, the Academy of Finland project 263457, the Center of Excellence ``Low Temperature Quantum Phenomena and Devices'' (project 250280) and the FP7  iQuOEMS project 323924.

\noindent{\bf Author Contributions}\\
P.L. and J.H. designed the sample. P.L. and P.H. were responsible for the experimental setup.  P.L. performed the experiment and analyzed the data. G.S.P. developed the theoretical model. P.L., P.H., and G.S.P wrote the paper, and the results were discussed with all authors.

\noindent{\bf Additional information}\\
{\bf Supplementary Information} accompanies this paper.
{\bf Competing financial interests:} The authors declare no competing financial interests.

\section{Methods}
{\bf Numerical evaluation of correlators}

The numerically simulated results presented in Fig. \ref{fig_spow} and Fig. \ref{fig_mim} were obtained using Gaussian (white) noise to represent the effect of the vacuum as input. Additionally, higher order reflections were included in Fig. \ref{fig_mim} by solving the Heisenberg-Langevin equation (HLE) in the rotating-wave approximation (RWA) $\dot{\tilde{a}} = \sum_{p=1,2}\alpha_{p} e^{-2i\Delta_{p}}\tilde{a}^{\dag} - \frac{\kappa}{2}\tilde{a} - \sqrt{\kappa}\tilde{a}_{\rm in}$. The presumption of white noise is accurate for narrow bandwidths where the spectrum of quantum noise is approximately uniform. In this way, higher-order reflections are included automatically in the results.

As seen from Fig. \ref{fig_mim}, the simple tripartite analytical solution of Eq. (\ref{finalsolution}), leading to correlators listed in Eqs. (5-9), agrees quite well with the results, except for the dark mode correlator $\langle\tilde{d}^\dagger \tilde{d}\rangle$. The disagreement can be traced to the assumption $\xi \ll \Delta_1, \Delta_2$, but the numerical simulations are not limited by this approximation.

{\bf Homodyne detection of correlations}

The output field of the resonator propagates through circulators and is amplified by a cryogenic low noise HEMT amplifier and by room-temperature microwave amplifiers. The quadratures are measured by standard homodyne methods, i.e. captured using an Anritsu MS2830A signal analyzer. Approximately 20 GB of data per sweep were collected and transferred to PC for later analysis. The single bins of the Fourier transformed data represent a bandwidth of 50 Hz and, therefore, spectral leakage due to used rectangular windowing is insignificant.

From the quadratures we can construct the complex field amplitudes, calculate their Fourier components around the central and extremal frequencies, in a bandwidth $BW$. We can write
\small
\begin{eqnarray}
\tilde{a}_{0} &=& \int_{-\infty}^{\infty} d\varsigma f_{(BW)}[\xi;\varsigma]\tilde{a}_{\rm out} [\varsigma], \label{ten}\\
\tilde{a}_{1,2} &=& \int_{-\infty}^{\infty} d\varsigma f_{(BW)}[2\Delta_{1,2}-\xi;\varsigma] \tilde{a}_{\rm out} [\varsigma ],\label{eleven}
\end{eqnarray}
\normalsize
where $f_{(BW)}[\xi;\varsigma]$ is a digital filtering function of width BW centered at the frequency $\xi$. Choosing $\int_{-\infty}^{\infty} d\varsigma|f_{(BW)}[\xi;\varsigma]|^2 = 1$ ensures that $[\tilde{a}_{0},\tilde{a}_{0}^\dagger] = 1$, $[\tilde{a}_{1},\tilde{a}_{1}^\dagger] = 1$, $[\tilde{a}_{2},\tilde{a}_{2}^\dagger] = 1$. A simple choice is $f_{(BW)}[\xi;\varsigma] = 1/\sqrt{BW}$ if $\varsigma$ is in the interval $(\xi-BW/2, \xi+BW/2)$, and 0 otherwise.

Because the added noise of the measurement chain is contained in $\tilde{a}_{\rm out}$ and it is uncorrelated, its contribution will vanish when calculating the correlations of $\tilde{a}_{0,1,2}$ and in the case of autocorrelation, it can be subtracted as a known constant. We can obtain then directly from Eqs. (\ref{ten}) and (\ref{eleven}) all the measured correlations, for example $\langle \tilde{a}_{0} \tilde{a}_{1} \rangle = \frac{1}{2} \cos \theta \exp (i\varphi_{1}) \sinh(2\lambda), $ etc.

{Also, given the definitions above, the correlations in the field at the output of the cavity are adimensional; in particular, $\langle \tilde{a}_0^\dagger \tilde{a}_0\rangle$, $\langle \tilde{a}_1^\dagger \tilde{a}_1\rangle$,  and $\langle \tilde{a}_2^\dagger \tilde{a}_2\rangle$ represent the flux of photons per bandwidth in the modes $\tilde{a}_0$, $\tilde{a}_1$, and $\tilde{a}_2$ respectively. This power is estimated from the increase of observed power due to pumping relative to the amplifier noise temperature.}

{\bf Pulsed parametric pumping}

In pulsed measurements, the pumping tones {(corresponding to 50 photon flux/Hz in continuous operation emerging from the cavity at the center frequency of 5 GHz)} were modulated using a pair of mixers (Marki M10220LA), driven by digital delay generators from Stanford Research System (DG535). The system provides an ON/OFF ratio of 1:100 and a rise time of 10 ns. The repetition rate was set to 250 kHz, which resulted in signals of 30\% when compared to continuous measurements. {With a pulse width of 1 $\mu$s, the repetition rate corresponds to a duty cycle of 1:3. The same method of simulating the Heisenberg-Langevin equations in time domain was employed to calculate numerically the response for pulsed pump tones.}

{\bf Data availability.} The authors declare that the data supporting the findings of this study are available within the article and its Supplementary Information files.\ \\


\begin{thebibliography}{10}
\expandafter\ifx\csname url\endcsname\relax
  \def\url#1{\texttt{#1}}\fi
\expandafter\ifx\csname urlprefix\endcsname\relax\def\urlprefix{URL }\fi
\providecommand{\bibinfo}[2]{#2}
\providecommand{\eprint}[2][]{\url{#2}}

\clearpage

\bibitem{purcelle}
\bibinfo{author}{Purcell, E. M.},
\newblock \bibinfo{title}{{Spontaneous Emission Probabilities at Radio Frequencies}}.
\newblock \emph{\bibinfo{journal}{Phys. Rev.}}
  \textbf{\bibinfo{volume}{69}}, \bibinfo{pages}{681}
  (\bibinfo{year}{1946}).
\newblock \urlprefix\url{http://journals.aps.org/pr/abstract/10.1103/PhysRev.69.674.2}

% 1
\bibitem{lamb}
\bibinfo{author}{Lamb, W.} \& \bibinfo{author}{Retherford, R.},
\newblock \bibinfo{title}{{Fine Structure of the Hydrogen Atom by a Microwave Method}}.
\newblock \emph{\bibinfo{journal}{Phys. Rev.}}
  \textbf{\bibinfo{volume}{72}}, \bibinfo{pages}{241-243}
  (\bibinfo{year}{1947}).
\newblock \urlprefix\url{http://dx.doi.org/10.1103/PhysRev.72.241}

% 2
\bibitem{schwinger}
\bibinfo{author}{Schwinger, J. S.},
\newblock \bibinfo{title}{{On gauge invariance and vacuum polarization}}
\newblock \emph{\bibinfo{journal}{Phys. Rev.}}
  \textbf{\bibinfo{volume}{82}}, \bibinfo{pages}{664-679}
  (\bibinfo{year}{1951}).
\newblock \urlprefix\url{http://dx.doi.org/10.1103/PhysRev.82.664}

% 3
\bibitem{hawking}
\bibinfo{author}{Hawking, S. W.},
\newblock \bibinfo{title}{{Black hole explosions?}}
\newblock \emph{\bibinfo{journal}{Nature}}
  \textbf{\bibinfo{volume}{248}}, \bibinfo{pages}{30-31}
  (\bibinfo{year}{1974}).
\newblock \urlprefix\url{http://dx.doi.org/10.1038/248030a0}

% 4
\bibitem{casimir}
\bibinfo{author}{Casimir, H. B. G.},
\newblock \bibinfo{title}{{On the attraction between two perfectly conducting plates}}
\newblock \emph{\bibinfo{journal}{Proc. K. Ned. Akad. Wet. B}}
  \textbf{\bibinfo{volume}{51}}, \bibinfo{pages}{793-795}
  (\bibinfo{year}{1948}).
%\newblock \urlprefix\url{http://www.mit.edu/~kardar/research/seminars/Casimir/Casimir1948.pdf}.

% 5
\bibitem{ussC}
G. S. Paraoanu, {\it The quantum vacuum}, in Romanian Studies in
Philosophy of Science (I. Parvu, G. Sandu, and I. Toader, eds.),
Boston Studies in the Philosophy and History of Science vol. 313,
Springer (2015) pp.181-197.
\newblock \urlprefix\url{http://link.springer.com/chapter/10.1007\%2F978-3-319-16655-1\%5F12}
%\bibinfo{author}{Paraoanu, G. S.}
%\newblock \emph{\bibinfo{title}{{The quantum vacuum}}}.
%\newblock Romanian Studies in Philosophy of Science (\bibinfo{publisher}{Springer International Switzerland},
%  \bibinfo{year}{2015}).

% 6
\bibitem{carloss}
\bibinfo{author}{Sab\'in, C.} \& \bibinfo{author}{Adesso, G.},
\newblock \bibinfo{title}{{Generation of quantum steering and interferometric power in the
Dynamical Casimir Effect}}.
\newblock \emph{\bibinfo{journal}{Phys. Rev. A}} \textbf{\bibinfo{volume}{92}},
  \bibinfo{pages}{042107} (\bibinfo{year}{2015}).
\newblock \urlprefix\url{http://dx.doi.org/10.1103/PhysRevA.92.042107}

% 7
\bibitem{ussA}
\bibinfo{author}{L\"ahteenm\"aki, P.},
\bibinfo{author}{Vesterinen, V.},
\bibinfo{author}{Hassel, J.}
\bibinfo{author}{Paraoanu, G.~S.},
\bibinfo{author}{Sepp\"a, H.},
\&
\bibinfo{author}{Hakonen, P.}
\newblock \bibinfo{title}{{Advanced Concepts in Josephson Junction Reflection Amplifiers}}.
\newblock \emph{\bibinfo{journal}{J. Low Temp. Phys.}} \textbf{\bibinfo{volume}{175}},
  \bibinfo{pages}{868-876} (\bibinfo{year}{2014}).
\newblock \urlprefix\url{http://dx.doi.org/10.1007/s10909-014-1170-0}

% 8
\bibitem{ussB}
\bibinfo{author}{Paraoanu, G. S.},
\newblock \bibinfo{title}{{Recent progress in quantum simulation using superconducting circuits}}.
\newblock \emph{\bibinfo{journal}{J. Low. Temp. Phys.}} \textbf{\bibinfo{volume}{175}},
  \bibinfo{pages}{633-654} (\bibinfo{year}{2014}).
\newblock \urlprefix\url{http://dx.doi.org/10.1007/s10909-014-1175-8}

% 9
\bibitem{nori2011}
\bibinfo{author}{You, J.~Q.} \& \bibinfo{author}{Nori, F.},
\newblock \bibinfo{title}{{Atomic physics and quantum optics using
  superconducting circuits.}}
\newblock \emph{\bibinfo{journal}{Nature}} \textbf{\bibinfo{volume}{474}},
  \bibinfo{pages}{589--597} (\bibinfo{year}{2011}).
\newblock \urlprefix\url{http://dx.doi.org/10.1038/nature10122}
%\newblock \eprint{arXiv:1202.1923v1}.

% 10
\bibitem{dpa2009}
\bibinfo{author}{Kamal, A.}, \bibinfo{author}{Marblestone, A.},
  \bibinfo{author}{Devoret, M.}
\newblock \bibinfo{title}{{Signal-to-pump back action and self-oscillation in double-pump Josephson parametric amplifier}}.
\newblock \emph{\bibinfo{journal}{Phys. Rev. B}} \textbf{\bibinfo{volume}{79}},
  \bibinfo{pages}{043804} (\bibinfo{year}{2009}).
\newblock \urlprefix\url{http://dx.doi.org/10.1103/PhysRevB.79.184301}

% 11
\bibitem{champara}
\bibinfo{author}{Simoen, M.},
%\bibinfo{author}{Chang, C. W. S.},
%\bibinfo{author}{Krantz, P.},
%\bibinfo{author}{Bylander, J.},
%\bibinfo{author}{Wustmann, W.},
%\bibinfo{author}{Shumeiko, V.},
%\bibinfo{author}{Delsing, P.} \&
%\bibinfo{author}{Wilson. C, M.},
\emph{et~al.}
\newblock \bibinfo{title}{{Characterization of a multimode coplanar waveguide parametric amplifier}}.
\newblock \emph{\bibinfo{journal}{J. Appl. Phys.}} \textbf{\bibinfo{volume}{118}},
  \bibinfo{pages}{154501} (\bibinfo{year}{2015}).
\newblock \urlprefix\url{http://dx.doi.org/10.1063/1.4933265}

% 12
\bibitem{eichlers}
\bibinfo{author}{Eichler, C.},
\bibinfo{author}{Salathe, Y.},
\bibinfo{author}{Mlynek, J.},
\bibinfo{author}{Schmidt, S.} \&
\bibinfo{author}{Wallraff, A.},
\bibinfo{author} \emph{et~al.}
\newblock \bibinfo{title}{{Quantum-Limited Amplification and Entanglement in Coupled Nonlinear Resonators}}.
\newblock \emph{\bibinfo{journal}{Phys. Rev. Lett.}} \textbf{\bibinfo{volume}{113}},
  \bibinfo{pages}{110502} (\bibinfo{year}{2014}).
\newblock \urlprefix\url{http://dx.doi.org/10.1103/PhysRevLett.113.110502}

% 13
\bibitem{yales}
\bibinfo{author}{Bergeal, N.},
%\bibinfo{author}{Schackert, F.},
%\bibinfo{author}{Metcalfe, M.},
%\bibinfo{author}{Vijay, R.},
%\bibinfo{author}{Manucharyan, V. E.},
%\bibinfo{author}{Frunzio, L.},
%\bibinfo{author}{Prober, D. E.},
%\bibinfo{author}{Schoelkopf, R. J.},
%\bibinfo{author}{Girvin, S. M.} \&
%\bibinfo{author}{Devoret, M. H.},
\bibinfo{author} \emph{et~al.},
\newblock \bibinfo{title}{{Phase-preserving amplification near the quantum limit with a Josephson ring modulator}}.
\newblock \emph{\bibinfo{journal}{Nature}} \textbf{\bibinfo{volume}{465}},
  \bibinfo{pages}{64-69} (\bibinfo{year}{2010}).
\newblock \urlprefix\url{http://dx.doi.org/10.1038/nature09035}

% 14
\bibitem{clerk2010}
\bibinfo{author}{Clerk, A.~A.}, \bibinfo{author}{Devoret, M.~H.},
  \bibinfo{author}{Girvin, S.~M.}, \bibinfo{author}{Marquardt, F.} \&
  \bibinfo{author}{Schoelkopf, R.~J.}
\newblock \bibinfo{title}{{Introduction to quantum noise, measurement, and
  amplification}}.
\newblock \emph{\bibinfo{journal}{Rev. Mod. Phys.}}
  \textbf{\bibinfo{volume}{82}}, \bibinfo{pages}{1155--1208}
  (\bibinfo{year}{2010}).
\newblock \urlprefix\url{http://link.aps.org/doi/10.1103/RevModPhys.82.1155}

% 15
\bibitem{yurke1988}
\bibinfo{author}{Yurke, B.},
%\bibinfo{author}{Kaminsky, P. G.},
%\bibinfo{author}{Miller, R. E.},
%\bibinfo{author}{Whittaker, E. A.},
%\bibinfo{author}{Smith, A. D.},
%\bibinfo{author}{Silver, A. H.} \&
%\bibinfo{author}{Simon, R. W.},
\emph{et~al.},
\newblock \bibinfo{title}{{Observation of 4.2-K equilibrium-noise squeezing via
  a Josephson-parametric amplifier}}.
\newblock \emph{\bibinfo{journal}{Phys. Rev. Lett.}}
  \textbf{\bibinfo{volume}{60}}, \bibinfo{pages}{764--767}
  (\bibinfo{year}{1988}).
\newblock \urlprefix\url{http://link.aps.org/doi/10.1103/PhysRevLett.60.764}

% 16
\bibitem{castellanos2008}
\bibinfo{author}{Castellanos-Beltran, M.~A.}, \bibinfo{author}{Irwin, K.~D.},
  \bibinfo{author}{Hilton, G.~C.}, \bibinfo{author}{Vale, L.~R.} \&
  \bibinfo{author}{Lehnert, K.~W.},
\newblock \bibinfo{title}{{Amplification and squeezing of quantum noise with a
  tunable Josephson metamaterial}}.
\newblock \emph{\bibinfo{journal}{Nat. Phys.}} \textbf{\bibinfo{volume}{4}},
  \bibinfo{pages}{929--931} (\bibinfo{year}{2008}).
\newblock \urlprefix\url{http://dx.doi.org/10.1038/nphys1090}

% 17
\bibitem{mallet2011}
\bibinfo{author}{Mallet, F.},
%\bibinfo{author}{Castellanos-Beltran, M. A.},
%\bibinfo{author}{Ku, H. S.},
%\bibinfo{author}{Glancy, S.},
%\bibinfo{author}{Knill, E.},
%\bibinfo{author}{Irwin, K. D.},
%\bibinfo{author}{Hilton, G. C.},
%\bibinfo{author}{Vale, L. R.} \&
%\bibinfo{author}{Lehnert, K. W.},
\emph{et~al.},
\newblock \bibinfo{title}{{Quantum State Tomography of an Itinerant Squeezed
  Microwave Field}}.
\newblock \emph{\bibinfo{journal}{Phys. Rev. Lett.}}
  \textbf{\bibinfo{volume}{106}}, \bibinfo{pages}{220502}
  (\bibinfo{year}{2011}).
\newblock
  \urlprefix\url{http://link.aps.org/doi/10.1103/PhysRevLett.106.220502}

% 18
\bibitem{eichler2011}
\bibinfo{author}{Eichler, C.},
%\bibinfo{author}{Bozyigit, D.},
%\bibinfo{author}{Lang, C.},
%\bibinfo{author}{Baur, M.},
%\bibinfo{author}{Steffen, L.},
%\bibinfo{author}{Fink, J. M.},
%\bibinfo{author}{Filipp, S.} \&
%\bibinfo{author}{Wallraff, A.},
\emph{et~al.},
\newblock \bibinfo{title}{{Observation of Two-Mode Squeezing in the Microwave
  Frequency Domain}}.
\newblock \emph{\bibinfo{journal}{Phys. Rev. Lett.}}
  \textbf{\bibinfo{volume}{107}}, \bibinfo{pages}{113601}
  (\bibinfo{year}{2011}).
\newblock
  \urlprefix\url{http://link.aps.org/doi/10.1103/PhysRevLett.107.113601}

% 19
\bibitem{wilson2011}
\bibinfo{author}{Wilson, C.~M.} \emph{et~al.}
\newblock \bibinfo{title}{{Observation of the dynamical Casimir effect in a
  superconducting circuit.}}
\newblock \emph{\bibinfo{journal}{Nature}} \textbf{\bibinfo{volume}{479}},
  \bibinfo{pages}{376-379} (\bibinfo{year}{2011}).
\newblock \urlprefix\url{http://dx.doi.org/10.1038/nature10561}

% 20
\bibitem{pasi2013}
\bibinfo{author}{L\"ahteenm\"aki, P.}, \bibinfo{author}{Paraoanu, G.~S.},
  \bibinfo{author}{Hassel, J.} \& \bibinfo{author}{Hakonen, P.~J.}
\newblock \bibinfo{title}{{Dynamical Casimir effect in a Josephson
  metamaterial}}.
\newblock \emph{\bibinfo{journal}{Proc. Natl. Acad. Sci.}}
  \textbf{\bibinfo{volume}{110}}, \bibinfo{pages}{4234--4238}
  (\bibinfo{year}{2013}).
\newblock \urlprefix\url{http://www.pnas.org/cgi/doi/10.1073/pnas.1212705110}
%\newblock \eprint{1111.5608}.

% 21
\bibitem{menzel2012}
\bibinfo{author}{Menzel, E.~P.} \emph{et~al.}
\newblock \bibinfo{title}{{Path Entanglement of Continuous-Variable Quantum
  Microwaves}}.
\newblock \emph{\bibinfo{journal}{Phys. Rev. Lett.}}
  \textbf{\bibinfo{volume}{109}}, \bibinfo{pages}{250502}
  (\bibinfo{year}{2012}).
\newblock
  \urlprefix\url{http://link.aps.org/doi/10.1103/PhysRevLett.109.250502}

% 22
\bibitem{ussD}
\bibinfo{author}{L\"ahteenm\"aki, P.}, \bibinfo{author}{Paraoanu, G.~S.},
  \bibinfo{author}{Hassel, J.} \& \bibinfo{author}{Hakonen, P.~J.},
\newblock \bibinfo{title}{{Photon Generation from Quantum Vacuum using a Josephson Metamaterial}}.
\newblock \emph{\bibinfo{journal}{PIERS in Taipei Proceedings}} \textbf{\bibinfo{volume}{151}},
  \bibinfo{pages}{} (\bibinfo{year}{2013}).
\newblock \urlprefix\url{https://piers.org/piersproceedings/piers2013TaipeiProc.php}

% 23
\bibitem{revmodph}
\bibinfo{author}{Nation, P.~D.}, \bibinfo{author}{Johansson, J.~R.},
  \bibinfo{author}{Blencowe, M.~P.}, \bibinfo{author}{Nori, Franco},
\newblock \bibinfo{title}{{Colloquium: Stimulating uncertainty: Amplifying the quantum vacuum with superconducting circuits}}.
\newblock \emph{\bibinfo{journal}{Rev. Mod. Phys.}}
  \textbf{\bibinfo{volume}{84}}, \bibinfo{pages}{1-24}
  (\bibinfo{year}{2012}).
\newblock \urlprefix\url{http://link.aps.org/doi/10.1103/RevModPhys.84.1}

% 24
\bibitem{chalmers-theory}
\bibinfo{author}{Bruschi, D.}, \bibinfo{author}{{\it et al.}}
\newblock \bibinfo{title}{{Towards universal quantum computation through
  relativistic motion}}.
\newblock \emph{\bibinfo{journal}{Sci. Rep.}} \bibinfo{pages}{18349}
(\bibinfo{year}{2016}).
\newblock
  \urlprefix\url{http://www.nature.com/articles/srep18349}

% 25
\bibitem{yuen1976}
\bibinfo{author}{Yuen, H.~P.},
\newblock \bibinfo{title}{{Two-photon coherent states of the radiation field}}.
\newblock \emph{\bibinfo{journal}{Phys. Rev. A}} \textbf{\bibinfo{volume}{13}},
  \bibinfo{pages}{2226--2243} (\bibinfo{year}{1976}).
\newblock
  \urlprefix\url{http://dx.doi.org/10.1103/PhysRevA.13.2226}

% 26
\bibitem{caves1985}
\bibinfo{author}{Caves, C.~M.} \& \bibinfo{author}{Schumaker, B.~L.}
\newblock \bibinfo{title}{{New formalism for 2-photon quantum optics. 1.
  Quadrature phases and squeezed states}}.
\newblock \emph{\bibinfo{journal}{Phys. Rev. A}} \textbf{\bibinfo{volume}{31}},
  \bibinfo{pages}{3068--3092} (\bibinfo{year}{1985}).
\newblock \urlprefix\url{http://dx.doi.org/10.1103/PhysRevA.31.3068}

% 27
\bibitem{qrng}
\bibinfo{author}{Symul, T.}, \bibinfo{author}{Assad, S. M.} \&
\bibinfo{author}{Lam, P. K.},
\newblock \bibinfo{title}{{Real time demonstration of high bitrate quantum random number generation with coherent laser light}}.
\newblock \emph{\bibinfo{journal}{Appl. Phys. Lett.}}
\textbf{\bibinfo{volume}{98}},
\bibinfo{pages}{231103}
(\bibinfo{year}{2011}).
\newblock \urlprefix\url{http://dx.doi.org/10.1063/1.3597793}

% 28
\bibitem{mandel1991}
\bibinfo{author}{Zou, X.~Y.}, \bibinfo{author}{Wang, L.~J.} \&
  \bibinfo{author}{Mandel, L.},
\newblock \bibinfo{title}{{Induced coherence and indistinguishability in
  optical interference}}.
\newblock \emph{\bibinfo{journal}{Phys. Rev. Lett.}}
  \textbf{\bibinfo{volume}{67}}, \bibinfo{pages}{318--321}
  (\bibinfo{year}{1991}).
\newblock \urlprefix\url{http://link.aps.org/doi/10.1103/PhysRevLett.67.318}

% 29
\bibitem{scully2000}
\bibinfo{author}{Kim, Y.-H.}, \bibinfo{author}{Yu, R.}, \bibinfo{author}{Kulik,
  S.~P.}, \bibinfo{author}{Shih, Y.} \& \bibinfo{author}{Scully, M.~O.},
\newblock \bibinfo{title}{{Delayed “Choice” Quantum Eraser}}.
\newblock \emph{\bibinfo{journal}{Phys. Rev. Lett.}}
  \textbf{\bibinfo{volume}{84}}, \bibinfo{pages}{1--5} (\bibinfo{year}{2000}).
\newblock \urlprefix\url{http://link.aps.org/doi/10.1103/PhysRevLett.84.1}

% 30
\bibitem{pappas2010}
\bibinfo{author}{Kelly, W.~R.} \emph{et~al.}
\newblock \bibinfo{title}{{Direct observation of coherent population trapping
  in a superconducting artificial atom}}.
\newblock \emph{\bibinfo{journal}{Phys. Rev. Lett.}}
  \textbf{\bibinfo{volume}{104}}, \bibinfo{pages}{163601} (\bibinfo{year}{2010}).
\newblock \urlprefix\url{http://dx.doi.org/10.1103/PhysRevLett.104.163601}

% 29
\bibitem{arimondo1996}
\bibinfo{author}{Arimondo, E.} \& \bibinfo{author}{Wolf, E.},
\newblock \bibinfo{title}{{Coherent population trapping in laser
  spectroscopy}}.
\newblock \emph{\bibinfo{journal}{Prog. Opt.}} \textbf{\bibinfo{volume}{35}},
  \bibinfo{pages}{257--354} (\bibinfo{year}{1996}).
%\newblock
%  \urlprefix\url{https://jila.colorado.edu/publications/coherent-population-trapping-laser-spectroscopy}.

% 30
\bibitem{vuleti2011}
\bibinfo{author}{Tanji-Suzuki, H.}, \bibinfo{author}{Chen, W.},
  \bibinfo{author}{Landig, R.}, \bibinfo{author}{Simon, J.} \&
  \bibinfo{author}{Vuletic, V.},
\newblock \bibinfo{title}{{Vacuum-Induced Transparency}}.
\newblock \emph{\bibinfo{journal}{Science}}
  \textbf{\bibinfo{volume}{333}}, \bibinfo{pages}{1266--1269}
  (\bibinfo{year}{2011}).
\newblock
  \urlprefix\url{http://www.sciencemag.org/cgi/doi/10.1126/science.1208066}

% 31
\bibitem{agarwal2013}
\bibinfo{author}{Agarwal, G.~S.},
\newblock \emph{\bibinfo{title}{{Quantum Optics}}}.
\newblock Quantum Optics (\bibinfo{publisher}{Cambridge University Press},
  \bibinfo{year}{2012}).
\newblock \urlprefix\url{https://books.google.fi/books?id=7KKw\%5FXIYaioC}

% 32
\bibitem{semba2014}
\bibinfo{author}{Zhu, X.} \emph{et~al.}
\newblock \bibinfo{title}{{Observation of dark states in a superconductor
  diamond quantum hybrid system.}}
\newblock \emph{\bibinfo{journal}{Nat. Commun.}} \textbf{\bibinfo{volume}{5}},
  \bibinfo{pages}{3424} (\bibinfo{year}{2014}).
\newblock \urlprefix\url{http://dx.doi.org/10.1038/ncomms4524}
%\newblock
%  \urlprefix\url{http://www.nature.com/ncomms/2014/140408/ncomms4524/full/ncomms4524.html}

% 33
\bibitem{kumar2015}
\bibinfo{author}{Kumar, K.~S.}, \bibinfo{author}{Veps\"al\"ainen, A.},
  \bibinfo{author}{Danilin, S.} \& \bibinfo{author}{Paraoanu, G.~S.},
\newblock \bibinfo{title}{{Stimulated Raman adiabatic passage in a three-level
  superconducting circuit}}.
\newblock \emph{\bibinfo{journal}{Nat. Commun.}} \textbf{\bibinfo{volume}{7}}, \bibinfo{pages}{10628} (\bibinfo{year}{2016}).
\newblock \urlprefix\url{http://dx.doi.org/10.1038/ncomms10628}
%\newblock
%  \urlprefix\url{http://www.nature.com/ncomms/2016/160223/ncomms10628/full/ncomms10628.html}

% 34
\bibitem{marquardt2007}
\bibinfo{author}{Marquardt, F.}, \bibinfo{author}{Chen, J.~P.},
  \bibinfo{author}{Clerk, A.~A.} \& \bibinfo{author}{Girvin, S.~M.},
\newblock \bibinfo{title}{{Quantum Theory of Cavity-Assisted Sideband Cooling
  of Mechanical Motion}}.
\newblock \emph{\bibinfo{journal}{Phys. Rev. Lett.}}
  \textbf{\bibinfo{volume}{99}}, \bibinfo{pages}{093902}
  (\bibinfo{year}{2007}).
\newblock
  \urlprefix\url{http://link.aps.org/doi/10.1103/PhysRevLett.99.093902}

% 35
\bibitem{zakka2011}
\bibinfo{author}{Zakka-Bajjani, E.} \emph{et~al.}
\newblock \bibinfo{title}{{Quantum superposition of a single microwave photon
  in two different ’colour’ states}}.
\newblock \emph{\bibinfo{journal}{Nat. Phys.}} \textbf{\bibinfo{volume}{7}},
  \bibinfo{pages}{599--603} (\bibinfo{year}{2011}).
\newblock \urlprefix\url{http://dx.doi.org/10.1038/nphys2035}

% 36
\bibitem{giedke}
\bibinfo{author}{Giedke, G.},
\bibinfo{author}{Kraus, B.},
\bibinfo{author}{Lewenstein, M.} \&
\bibinfo{author}{Cirac, J. I.},
\newblock \bibinfo{title}{{Suppression of arbitrary internal coupling in a quantum register}}.
\newblock \emph{\bibinfo{journal}{Phys. Rev. A}} \textbf{\bibinfo{volume}{64}},
  \bibinfo{pages}{052301} (\bibinfo{year}{2001}).
\newblock \urlprefix\url{http://dx.doi.org/10.1103/PhysRevA.64.052301}

% 37
\bibitem{adesso}
\bibinfo{author}{Adesso, G.},
\bibinfo{author}{Serafini, A.}, \&
\bibinfo{author}{Illuminati, F.},
\newblock \bibinfo{title}{{Multipartite entanglement in three-mode Gaussian states of continuous-variable systems: Quantification, sharing structure, and decoherence}}.
\newblock \emph{\bibinfo{journal}{Phys. Rev. A}} \textbf{\bibinfo{volume}{73}},
  \bibinfo{pages}{032345} (\bibinfo{year}{2006}).
\newblock \urlprefix\url{http://dx.doi.org/10.1103/PhysRevA.73.032345}

% 38
\bibitem{serafini}
\bibinfo{author}{Serafini, A.},
\bibinfo{author}{Adesso, G.}, \&
\bibinfo{author}{Illuminati, F.},
\newblock \bibinfo{title}{{Unitarily localizable entanglement of Gaussian states}}.
\newblock \emph{\bibinfo{journal}{Phys. Rev. A}}
\textbf{\bibinfo{volume}{71}},
\bibinfo{pages}{032349} (\bibinfo{year}{2005}).
\newblock \urlprefix\url{http://dx.doi.org/10.1103/PhysRevA.71.032349}

% 39
\bibitem{gadesso}
\bibinfo{author}{Adesso, G.},
\bibinfo{author}{Serafini, A.}, \&
\bibinfo{author}{Illuminati, F.},
\newblock \bibinfo{title}{{Quantification and Scaling of Multipartite Entanglement in Continuous Variable Systems}}.
\newblock \emph{\bibinfo{journal}{Phys. Rev. Lett.}}
\textbf{\bibinfo{volume}{93}},
\bibinfo{pages}{220504} (\bibinfo{year}{2004}).
\newblock \urlprefix\url{http://dx.doi.org/10.1103/PhysRevLett.93.220504}

% 40
\bibitem{dur}
\bibinfo{author}{D\"ur, W.},
\bibinfo{author}{Vidal, G.}, \&
\bibinfo{author}{Cirac, J. I.},
\newblock \bibinfo{title}{{Three qubits can be entangled in two inequivalent ways}}.
\newblock \emph{\bibinfo{journal}{Phys. Rev. A}}
\textbf{\bibinfo{volume}{62}},
\bibinfo{pages}{062314} (\bibinfo{year}{2000}).
\newblock \urlprefix\url{http://dx.doi.org/10.1103/PhysRevA.62.062314}

% 41
\bibitem{loock2001}
\bibinfo{author}{van Loock, P.} \& \bibinfo{author}{Braunstein, S. L.},
\newblock \bibinfo{title}{{Telecloning of Continuous Quantum Variables}}.
\newblock \emph{\bibinfo{journal}{Phys. Rev. Lett.}} \textbf{\bibinfo{volume}{87}},
  \bibinfo{pages}{247901} (\bibinfo{year}{2001}).
\newblock \urlprefix\url{http://dx.doi.org/10.1103/PhysRevLett.87.247901}

% 42
\bibitem{adesso2007}
\bibinfo{author}{Adesso, G.}, \bibinfo{author}{Serafini, A.} \&
\bibinfo{author}{Illuminati, F.},
\newblock \bibinfo{title}{{Optical state engineering, quantum communication, and robustness of entanglement promiscuity in three-mode Gaussian states}}.
\newblock \emph{\bibinfo{journal}{New J. Phys.}} \textbf{\bibinfo{volume}{9}},
  \bibinfo{pages}{60} (\bibinfo{year}{2007}).
\newblock \urlprefix\url{http://dx.doi.org/10.1088/1367-2630/9/3/060}

% 43
\bibitem{realism}
\bibinfo{author}{Paraoanu, G. S.},
\newblock \bibinfo{title}{{Realism and single-quanta nonlocality}}.
\newblock \emph{\bibinfo{journal}{Found. Phys.}}
\textbf{\bibinfo{volume}{41}},
\bibinfo{pages}{734} (\bibinfo{year}{2011}).
\newblock \urlprefix\url{http://dx.doi.org/10.1007/s10701-010-9513-4}

% 44
\bibitem{thispaper}
\bibinfo{author}{Heaney, L.},
\bibinfo{author}{Cabello, A.},
\bibinfo{author}{Santos, M. F.}, \&
\bibinfo{author}{Vedral, V.},
\newblock \bibinfo{title}{{Extreme nonlocality with one photon}}.
\newblock \emph{\bibinfo{journal}{New J. Phys.}}
\textbf{\bibinfo{volume}{13}},
\bibinfo{pages}{053054} (\bibinfo{year}{2011}).
\newblock \urlprefix\url{http://iopscience.iop.org/article/10.1088/1367-2630/13/5/053054}

% 45
\bibitem{aaronson}
\bibinfo{author}{Aaronson, S.} \&
\bibinfo{author}{Arkhipov, A.},
\newblock \bibinfo{title}{{The computational complexity of linear optics}}.
\newblock \emph{\bibinfo{journal}{Proceedings of the forty-third annual ACM symposium on Theory of computing}}
\textbf{\bibinfo{volume}{}},
\bibinfo{pages}{333-342} (\bibinfo{year}{2011}).
\newblock \urlprefix\url{http://dx.doi.org/10.1145/1993636.1993682}

% 46
\bibitem{pfister2014}
\bibinfo{author}{Chen, M.}, \bibinfo{author}{Menicucci, N.~C.} \&
  \bibinfo{author}{Pfister, O.},
\newblock \bibinfo{title}{{Experimental Realization of Multipartite
  Entanglement of 60 Modes of a Quantum Optical Frequency Comb}}.
\newblock \emph{\bibinfo{journal}{Phys. Rev. Lett.}}
  \textbf{\bibinfo{volume}{112}}, \bibinfo{pages}{120505}
  (\bibinfo{year}{2014}).
\newblock
  \urlprefix\url{http://link.aps.org/doi/10.1103/PhysRevLett.112.120505}

% 47
\bibitem{hans}
\bibinfo{author}{Raussendorf, R.} \& \bibinfo{author}{Briegel, H. J.},
\newblock \bibinfo{title}{{A One-Way Quantum Computer}}.
\newblock \emph{\bibinfo{journal}{Phys. Rev. Lett.}}
  \textbf{\bibinfo{volume}{86}}, \bibinfo{pages}{5188}
  (\bibinfo{year}{2001}).
\newblock
  \urlprefix\url{http://link.aps.org/doi/10.1103/PhysRevLett.86.5188}

% 48
\bibitem{seppaVTT}
\bibinfo{author}{Kiviranta, M.},
\emph{et~al.},
%, Brandel, O., Gr\"onberg, L., Kunert, J., Linzen, S., Beev, N. \& Prunnila}, M.},
\newblock \bibinfo{title}{{Multilayer fabrication process for josephson
  junction circuits cross-compatible over two foundries}}.
\newblock \emph{\bibinfo{journal}{IEEE Trans. Appl. Supercond.}}
  \textbf{\bibinfo{volume}}{99}, \bibinfo{pages}{1-5} (\bibinfo{year}{2016}).
\newblock
  \urlprefix\url{http://dx.doi.org/10.1109/TASC.2016.2544821}

% 49
\bibitem{zeilinger2009}
\bibinfo{author}{Ramelow, S.}, \bibinfo{author}{Ratschbacher, L.},
  \bibinfo{author}{Fedrizzi, A.}, \bibinfo{author}{Langford, N.~K.} \&
  \bibinfo{author}{Zeilinger, A.},
\newblock \bibinfo{title}{{Discrete tunable color entanglement}}.
\newblock \emph{\bibinfo{journal}{Phys. Rev. Lett.}}
  \textbf{\bibinfo{volume}{103}}, \bibinfo{pages}{253601} (\bibinfo{year}{2009}).
\newblock
  \urlprefix\url{http://dx.doi.org/10.1103/PhysRevLett.103.253601}

% 50
\bibitem{zeilinger2014}
\bibinfo{author}{Lemos, G.~B.}, \emph{et~al.},
\newblock \bibinfo{title}{{Quantum imaging with undetected photons}}.
\newblock \emph{\bibinfo{journal}{Nature}} \textbf{\bibinfo{volume}{512}},
  \bibinfo{pages}{409--412} (\bibinfo{year}{2014}).
\newblock \urlprefix\url{http://dx.doi.org/10.1038/nature13586}

% 51
\bibitem{review-delayed}
\bibinfo{author}{Ma, X.-S.}, \bibinfo{author}{Kofler, J.} \&
  \bibinfo{author}{Zeilinger, A.},
\newblock \bibinfo{title}{{Delayed-choice gedanken experiments and their
  realizations}}.
\newblock \emph{\bibinfo{journal}{Rev. Mod. Phys}}
  \textbf{\bibinfo{volume}{88}} \bibinfo{pages}{015005} (\bibinfo{year}{2016}).
\newblock \urlprefix\url{http://dx.doi.org/10.1103/RevModPhys.88.015005}

% 52
\bibitem{haroche} Bertet, P., {\it et al.} A complementarity experiment with an interferometer at the quantum-classical boundary, {\it Nature} {\bf 411}, 166-170 (2001).
\newblock \urlprefix\url{http://dx.doi.org/10.1038/35075517}

% 53
\bibitem{loock2005}
\bibinfo{author}{Braunstein, S.~L.} \& \bibinfo{author}{van Loock, P.},
\newblock \bibinfo{title}{{Quantum information with continuous variables}}.
\newblock \emph{\bibinfo{journal}{Rev. Mod. Phys.}}
  \textbf{\bibinfo{volume}{77}}, \bibinfo{pages}{513--577}
  (\bibinfo{year}{2005}).
\newblock \urlprefix\url{http://link.aps.org/doi/10.1103/RevModPhys.77.513}
%\newblock \eprint{0410100}.

% 54
\bibitem{DaSilva2010}
\bibinfo{author}{da~Silva, M.~P.}, \bibinfo{author}{Bozyigit, D.},
  \bibinfo{author}{Wallraff, A.} \& \bibinfo{author}{Blais, A.},
\newblock \bibinfo{title}{{Schemes for the observation of photon correlation
  functions in circuit QED with linear detectors}}.
\newblock \emph{\bibinfo{journal}{Phys. Rev. A}} \textbf{\bibinfo{volume}{82}},
  \bibinfo{pages}{043804} (\bibinfo{year}{2010}).
\newblock \urlprefix\url{http://link.aps.org/doi/10.1103/PhysRevA.82.043804}

\end{thebibliography}
\end{document}